# Electronic Structure of Mott Insulators Studied by Inelastic X-ray Scattering

M. Z. Hasan,[1][*] E. D. Isaacs,[2] Z-X. Shen,[1] L. L. Miller,[3] K. Tsutsui,[4] T. Tohyama,[4] S. Maekawa[4]

1. Department of Applied Physics, Physics and Stanford Synchrotron Radiation Laboratory, Stanford Linear Accelerator Center (SLAC), Stanford University, Stanford, CA 94305
2. Bell Laboratories, Lucent Technologies, Murray Hill, NJ 07974
3. Department of Physics & Ames Laboratory, Iowa State University, Ames, IA 50011
4. Institute of Materials Research, Tohoku University, Sendai 980-8577, Japan

* To whom correspondence should be addressed. E-mail : mzhasan@Stanford.edu

The electronic structure of Mott insulators continues to be a major unsolved problem in physics despite more than half-century of intense research efforts. Well-developed momentum-resolved spectroscopies such as photoemission or neutron scattering cannot probe the full Mott gap. We report observation of dispersive charge excitations across the Mott gap in a high $T_c$ parent cuprate $Ca_2CuO_2Cl_2$ using high resolution resonant inelastic x-ray scattering shedding light on the anisotropy of Mott-gap. The results provide direct support for the Hubbard model to describe charge excitations across the Mott gap.

The discovery of high temperature superconductivity and colossal magnetoresistance in doped transition metal oxides has led to the extensive research interests in Mott insulators. Such oxides are characterized by large onsite Coulomb interaction and the consequent low-temperature insulating state characterized by a charge-excitation gap known as the Mott-gap. The gap is either set by the Coulomb interaction $U$ or the charge-transfer energy $\Delta$ (energy to remove an electron from oxygen orbital and put it on the copper site) depending on which one is lower *(1-4)*. Angle-resolved photoemission (ARPES) which probes only the occupied electronic states has been remarkably successful in characterizing the electronic structure of cuprate based insulators *(5-8)*. Little is known about the momentum (k-) resolved electronic structure of the unoccupied band which is a major barrier for a coherent understanding of the nature of the Mott gap and its related insulating state. In addition, knowledge of the unoccupied upper Hubbard band is essential to understand the physics of *n*-type (electron doped) superconductors as the doped electrons occupy the upper Hubbard band. Among the standard probes of condensed matter systems (that allow momentum-resolved studies), neutrons do not couple to electron's charge density and (thermal) neutron energy is too low to reach the Mott-edge. No **k**-resolved inverse photoemission (Inv-ARPES) study is available because of problems associated with sample charging as well as the lack of required energy resolution. Inelastic electron scattering which is known as the electron-energy-loss spectroscopy (EELS) is possible and measures electronic excitations from the occupied to the unoccupied bands. Though very useful as a technique, EELS requires extensive sample preparation and the spectra need to be corrected for multiple scattering effects in order to extract useful information *(9)*.

Inelastic x-ray scattering is a natural and powerful probe of electronic excitations in condensed matter systems. It has the potential to fill an important gap of knowledge by developing a good understanding of the bulk electronic structure of correlated electron systems. Inelastic scattering of x-ray photons covers a fairly wide kinematic range in energy and momentum space and the photons directly couple to the electronic charge (and other electronic degrees of freedom like orbitals and spins). However, as x-ray photons are highly absorbed in high-$Z$ materials, applications of the technique have been mostly limited to low-$Z$ systems *(10-13)*. Several recent studies, both experimental results and theoretical/numerical investigations have shown that by tuning the incident photon energy near an x-ray absorption edge a Raman-like effect could be measured with nonzero momentum transfer, despite the high absorption cross-section, through the large resonant enhancement which eventually dominates the overall cross-section *(14-19)*. An inelastically scattered x-ray photon can probe the full charge-gap in a Mott insulator through the creation of a hole in the occupied band and promoting an electron across the gap to the unoccupied band with a finite (tunable) momemtum transferred into the system. A recent resonant inelastic x-ray scattering (RIXS) study *(17)* has reported such observation of a low-energy charge-transfer gap near 2.5 eV at finite-$q$ along the <210> direction in a parent cuprate $Sr_2CuO_2Cl_2$ whose electronic structure is quite similar to $Ca_2CuO_2Cl_2$ (6,7,8). A similar excitation band has also been reported at finite-$q$ in electron-energy-loss spectroscopy (9).

So far, however, no $q$-resolved x-ray study has been reported along the high symmetry directions. In this paper, we report the first direct bulk (~ 10-20 micron) measurement of the electronic excitations in the cuprates up to 8 eV energy-loss along two high symmetry directions (the Cu-O bond direction <100> and 45 degrees to the bond-direction <110>) of the $CuO_2$ plane. $Ca_2CuO_2Cl_2$ was chosen for its relatively low x-ray absorption and high crystalline quality (20). Experiments were performed on the X-21 triple-axis spectrometer at the National Synchrotron Light Source which had been described in detail elsewhere (14,15). With the incident photon energy fixed on resonance near the Cu K-edge at $E_0$ = 8.996 KeV, energy analysis was performed by scanning the analyzer. The overall energy resolution of the spectrometer was 440 meV as determined by the energy width of the elastic peak. The background was measured by keeping track of scattering intensities on the energy gain side and averaged about 1-2 counts per minute.

Inelastic x-ray scattering spectra in the parent cuprate insulator $Ca_2CuO_2Cl_2$ with the incident energy near the Cu $K_\alpha$ absorption edge ($E_0$=8.996 KeV) were measured for a range of momentum transfers. The spectra are presented in Fig.1A along the <110> and in Fig.1B along the <100> direction. The spectra along the <110>-direction are taken in the second Brillouin zone whereas the spectra along the <100> direction are in the third Brillouin zone. Selection of different Brillouin zones was due to various kinematic limitations with the spectrometer configuration. Each spectrum exhibits two inelastic features – one broad peak around 5.8 eV and a weak feature which disperses with $q$ ranging from 2.5 to 3.8 eV. The broad feature centered around 5.8 eV, reported earlier along different directions <001> and <210> (15,17) changes shape with change of $q$ (and polarization) but does not show any significant dispersion in either the <100> or <110> direction. The lack of dispersion establishes the highly local character of this excitation. Based on electronic structure calculations the 5.8-eV feature is believed to be excitations to the unoccupied copper orbitals with antibonding character (15,21). More precisely, it has been identified as corresponding to a transition from the ground state to a high-energy excited state composed of symmetric contributions of a central Cu-$3d_{x2-y2}$ orbital and the surrounding O-$2p_\sigma$ orbitals (17,22). In contrast, the position of the lower energy feature, changes significantly with $q$ from (0,0) (measured near $(2\pi,2\pi)$) to $(\pi,\pi)$ as seen in Fig.1A. This feature first emerges clearly from the tail of the elastic peak near $(\pi/2,\pi/2)$ and disperses upward about 1.35 eV monotonically compared to its position near (0,0). At q = $(\pi,\pi)$, where the low energy excitation has dispersed to 3.8 eV, we note a shoulder of the elastic peak at a lower energy-loss near 2.8 eV. With the scattering wave vector, $q$, along the <100>-direction (Fig.1B), the low-energy feature appears as a shoulder on the elastic peak in the raw data for $q$ < $(\pi/2,0)$. From $(\pi/2, 0)$ to $(\pi, 0)$ it disperses upward by only about 0.56 eV. In order to make our estimates of the positions of the low energy inelastic feature more reliable, particularly for low $q$, the elastic peak was removed from the raw data by fitting. Because the subtraction of the fit from the data is imperfect, we do not extract any reliable information about the electronic structure within 1 eV of the elastic peak. Selected spectra with the elastic scattering removed are shown in Fig. 2A along <110> and Fig. 2B along <100>. Along <110> dispersion of the low-energy feature is monotonically upward. The spectrum for $q$ =$(\pi,\pi)$ shows [Fig. 2A topmost spectrum] an even lower energy component near 2.8 eV in addition to the 3.8 eV feature discussed earlier. Along <100> this excitation (the feature's center of gravity) is nearly non-dispersive for $q$ < $(\pi/2,0)$ but disperses weakly and monotonically upward for $q$ > $(\pi/2,0)$. The center of gravity of the low-energy inelastic feature are shown in Fig. 3A and 3B as a function of $q$ along the <110> and <100> directions, respectively (denoted by open circles). Relative excitation energies are plotted referenced to the energy at $q$ = (0,0) along each direction. Due to the doublet nature of the excitation at $q$ =$(\pi,\pi)$, in addition to plotting the center of gravity, we plot the energy position of two separable features. A $q$-space map of charge excitations across the Mott gap is shown in Fig. 3C which reveals the overall shape of the dispersion surface.

We interpret the low-energy feature in the data as the observation of $q$-resolved excitations across the effective Mott-gap - from the occupied band consisting (23) of Cu $3d_{x2-y2}$ and O-$2p_\sigma$ orbitals (Zhang-Rice band (ZRB)) to the unoccupied upper Hubbard band (UHB) along high symmetry directions. The effective Mott (or charge-transfer) excitation is shown in an energy level schematic ($k$-integrated) for this cuprate insulator (Fig. 4A). The RIXS process shakes up the valence electron system and creates a particle-hole pair across the gap which carries energy and momentum. This pair propagates in a background of antiferromagnetically ordered lattice. In a local picture, the created hole forms a Zhang-Rice singlet *(23)* and an electron is excited onto the neighboring Cu site occupying the UHB. Fig. 4B shows a schematic of a particle-hole pair excitation in the $CuO_2$ plane (the arrows denote the spins of holes). Strong antiferromagnetic correlation (unusually large superexchange coupling, J ~ 0.12 eV *(6)*) of the system plays a key role in determining the anisotropic propagation of the particle-hole excitations. In addition, the upper Hubbard band in cuprates is believed to be strikingly different from the conduction band of a normal semiconductor or a band insulator as revealed by X-ray absorption spectroscopy *(24)*. We note that in the absence of strong electron-electron correlation such as in case of the band insulators a convolution of the 'valence band' and the 'conduction band' is sufficient to understand their RIXS spectra *(25)* whereas for a Mott insulator such an approach is not sufficient because the gap does not originate from band effects but rather from the strong correlation effects so the particle-hole excitations of the Mott system can not be treated as independent excitations of single-particle states *(18,24)*. Consideration of two-particle Green's functions is necessary to interpret RIXS spectra. In this sense, RIXS would provide more information than combined ARPES and Inv-ARPES. A model based rigorous calculation is necessary to elucidate the origin and nature of the **q**-resolved charge excitations in a strongly correlated quantum system.

Several approaches have, so far, been proposed to interpret the RIXS process *(15-17,21)* in cuprates. The clear dispersions observed along the high symmetry directions in our experimental data make it possible to perform a direct quantitative comparison to the results of specific many-body model calculations. We consider the Hubbard model with long-range hopping where the lower Hubbard band (LHB) is regarded as the Zhang-Rice band and the Hamiltonian ($t$-$t'$-$t''$-$U$ model) is given by

$$H_{3d} = -t \sum_{<i,j>,\sigma}(d_{i,\sigma}^{\dagger}d_{j,\sigma}+h.c.) - t' \sum_{<i,j>',\sigma}(d_{i,\sigma}^{\dagger}d_{j,\sigma}+h.c.) - t'' \sum_{<i,j>'',\sigma}(d_{i,\sigma}^{\dagger}d_{j,\sigma}+h.c.) + U \sum_i n_{i,\uparrow}^d n_{i,\downarrow}^d$$

where $d_{i,\sigma}^{\dagger}$ is the creation operator of 3$d$ electron with spin σ at site i, $n_{i,\sigma}^d = d_{i,\sigma}^{\dagger}d_{i,\sigma}$, and $n_i^d = n_{i,\uparrow}^d + n_{i,\downarrow}^d$. The summations $<i,j>$, $<i,j>'$, and $<i,j>''$ run over the first, second, and third nearest neighbor pairs, respectively. The values of the parameters are taken from fits to ARPES data on $Sr_2CuO_2Cl_2$ *(8)* to be $t = 0.35$ eV, $U/t = 10$, $t'/t = -0.34$, and $t''/t = 0.23$ where $t, t'$ and $t''$ are the first, second and third nearest neighbor hopping and $U$ is the on-site Coulomb interaction *(8,18)*. In the intermediate states of Cu K-edge RIXS process, 3$d$ electrons interact with a 1$s$ core hole created by the dipole transition of a 1$s$ electron to 4$p$ orbital due to the absorption of an incident photon with energy $\omega_i$ and momentum $\mathbf{K_i}$. This interaction Hamiltonian is written as: $H_{1s-3d} = -V_c \sum_{i,\sigma,\sigma'} n_{i,\sigma}^d n_{i,\sigma'}^s$; where $n_{i,\sigma}^s$ is the number operator of 1$s$ core hole with spin σ at site i, and $V_c$ is taken to be positive. This interaction causes excitations of the 3$d$ electrons across the gap. The photo-excited 4$p$ electron is assumed to go into the bottom of the 4$p$ band and not to interact with either the 3$d$ electrons or the 1$s$ core hole due to delocalized nature of the 4$p$ bands. In the final state, the 4$p$ electron goes back to the 1$s$ orbital emitting a photon with energy $\omega_f$ and momentum $\mathbf{K_f}$. It is also assumed that the momentum transfer, $\mathbf{q} = \mathbf{K_i}-\mathbf{K_f}$ is conserved within the electronic degrees of freedom. In this scenario, the RIXS spectrum is given by *(18)*:

$$I(q,\Delta\omega) = \sum_\alpha |<\alpha|\sum_\sigma s_{k0-Kf,\sigma} p_{k0,\sigma} (H+\varepsilon_{1s-4p}-E_0-\omega_i-i\Gamma)^{-1} p_{k0,\sigma}^{\dagger} s_{k0-Ki,\sigma}^{\dagger} |0>|^2 \delta(\Delta\omega-E_\alpha+E_0),$$

where $H = H_{3d} + H_{1s-3d}$, $\mathbf{q} = \mathbf{K_i} - \mathbf{K_f}$, $\Delta\omega = \omega_i - \omega_f$, $s_{k,\sigma}^{\dagger}$ ($p_{k,\sigma}^{\dagger}$) is the creation operator of the 1$s$ core hole (4$p$ electron) with momentum **k** and spin σ, |0> is the ground state of the half-filled system with energy $E_0$, |α> is the final state of the RIXS process with energy $E_\alpha$, $\Gamma$ is the inverse of the relaxation time in the intermediate state, and $\varepsilon_{1s-4p}$ is the energy difference between the 1$s$ level and the bottom of the 4$p$ band. The values of the parameters are set to be $V_c/t = 15$ and $\Gamma/t = 1$ *(18)*.

Fig. 2C and 2D show the RIXS spectra calculated on a 4x4-site cluster. Along the <110> direction, the energy position of the features increases monotonically with increasing **q**. In contrast to the <110> direction, the spectral threshold along the <100> direction decreases in energy from $\mathbf{q} = (0,0)$ to $(\pi/2,0)$, and then increases toward $\mathbf{q} = (\pi,0)$. These features are overall consistent with the low energy features of the experimental spectra (Fig. 2A and 2B). For more quantitative comparison between experiment and model, we plot the momentum dependence of the center of gravity of the experimental and numerical results (Fig. 3A and 3B). Along the <110> direction, the overall trend of the center of gravity is similar between the experiment and the model. For the $\mathbf{q} = (\pi,\pi)$ spectrum we plot two centers of gravity due to two separable low-energy features [Fig. 2A, top spectrum]. Note that the theory curve at **q** = (π,π) also contains two features that are most separated in energy [Fig. 2C, top spectrum]. The lower energy peak near 2.8 eV in the 'doublet' seen for **q** = (π,π) spectrum [Fig. 2A, top spectrum] is likely to be of excitonic origin indicating the possibility of intersite Coulomb correlations. Along the <100> direction, the correspondence of the centers of the gravity between experiment and theory is reasonably good too. The flatness in energy dispersion from **q** = (0,0) to (π/2,0) [Fig. 3B] is likely to be due to the fact that in *t-t'-t''-U* model, that is, in the presence of higher-order hopping, the single-particle spectral function ( $A(\mathbf{k},\omega)$ ) for the unoccupied states (upper Hubbard band) has a minimum near **k** = (π, 0) unlike the occupied Zhang-Rice band's maximum which is near **k** = (π/2,π/2) *(6,7,9,22)*. Such flatness is not expected in *t-U* model *(18,26)*. So these results suggest that the unoccupied band has a different symmetry compared to the occupied band. This disparity between the occupied and the unoccupied bands can potentially shed light in understanding the origin of different behaviors of *n*-type and *p*-type cuprate superconductors. It is interesting to note that at this level of experimental resolution it is not necessary to consider different values for the hoppings between the Zhang-Rice band and the upper Hubbard band to describe these results. Further higher resolution work is definitely necessary to resolve the details of many of these issues. We conclude that these results, in general, indicate that the **q**-resolved charge excitations across the Mott gap can be described within the general framework of the standard Hubbard model.

These experimental results demonstrate the power and versatility of such x-ray spectroscopy in addressing some fundamental issues of charge excitations in Mott-insulators that are not directly possible by other well-developed spectroscopic techniques. With the availability of higher resolution and brighter synchrotron sources, a more detailed study (such as the possibility of the broken particle-hole symmetry as hinted here, the magnitude of intersite Coulomb coupling, the possible existence of RVB-like *(27)* correlations etc.) of Mott insulators would be possible. Feasibility of such experiments points to the opening of a new frontier of correlated system research.

We gratefully acknowledge P. Abbamonte, S. Ahmed, P. W. Anderson, A. I. Bienenstock, C. A. Burns, S. Doniach, Z. Hussain, C. C. Kao, S. A. Kivelson, M. V. Klein, R. B. Laughlin, V. Oganesyan, P. M. Platzman, F. Ronning, G. A. Sawatzky, J. Zaanen, S. C. Zhang and X. J. Zhou for useful suggestions. The work at NSLS/Brookhaven was jointly supported by Bell-Labs, Lucent Technologies and DOE/BES (Materials Sciences Division). The theoretical/numerical work was supported by Priority-Areas Grants from the Ministry of Education, Science, Culture and Sport of Japan, CREST, and NEDO. Computations were carried out in ISSP (University of Tokyo), IMR (Tohoku University) and Tohoku University.

# Fig. 1 (Experimental Spectra)

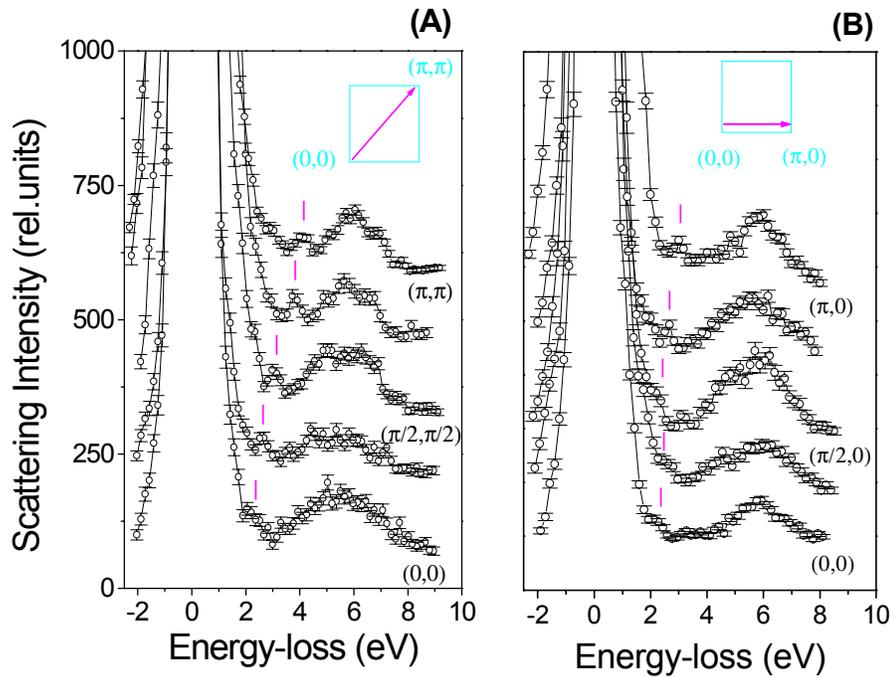

**Hasan *et.al.*, Science 288, 1811 (2000)**

Fig. 1: (**Experimental spectra**) **q**-dependence of inelastic x-ray scattering spectra near Cu K-edge are shown along two high symmetry directions in $Ca_2CuO_2Cl_2$ : (A) Scattering along the <110>-direction from equivalent-(0,0) to $(\pi,\pi)$ (B) Scattering along the <100>-direction from equivalent-(0,0) to $(\pi,0)$. Incident photon energy $E_o$ = 8.996 eV.

# Fig. 2 (Extracted Data and Model Comparison)

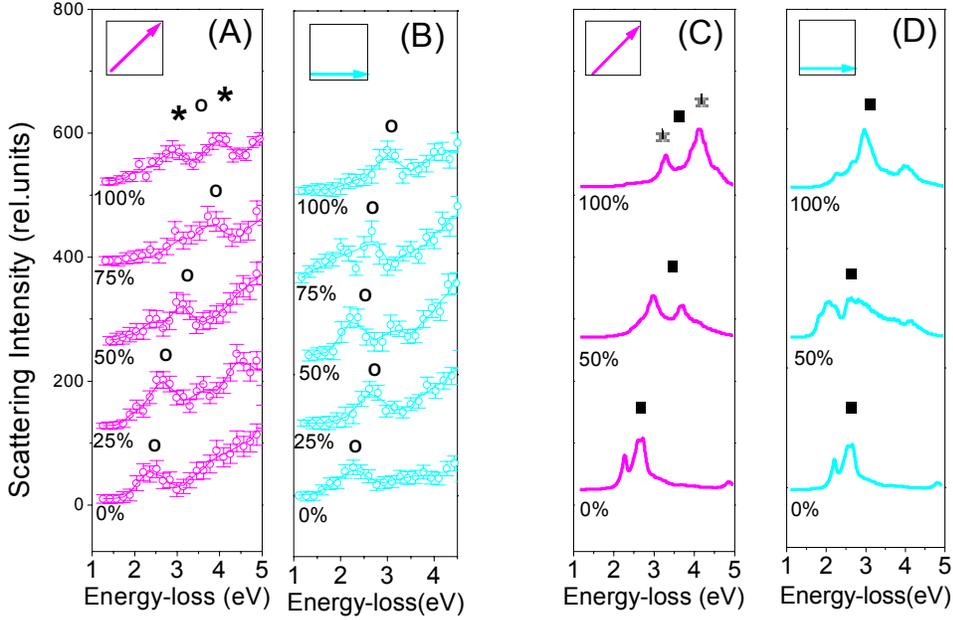

**Hasan *et.al.*, Science 288, 1811 (2000)**

Fig. 2: (**Extracted data and model comparison**) Momentum dependence of the RIXS spectra are compared with Hubbard model calculations : the experimental spectra (with elastic scattering removed) are shown along the <110> direction (A) and the <100> direction (B) and the calculated ones (low-energy component only) are along the <110> direction (C) and the <100> direction (D). The parameter values for (C) and (D) are $t$ = 0.35 eV, $U/t$ = 10, $t'/t$ = -0.34, and $t''/t$ = 0.23, $V_c/t$ = 15 *(8,18)*. The centers of gravity of spectral weights are indicated by the open circles for the experimental data and closed squares for the $t$-$t'$-$t''$-$U$ Hubbard model calculations. The percent numbers show the normalized **q** co-ordinates (locations inside the Brillouin zone) and the inset in each panel shows the direction of the scattering wave vector, **q** (the momentum transfer).

# Fig. 3 ( Dispersion Plots)

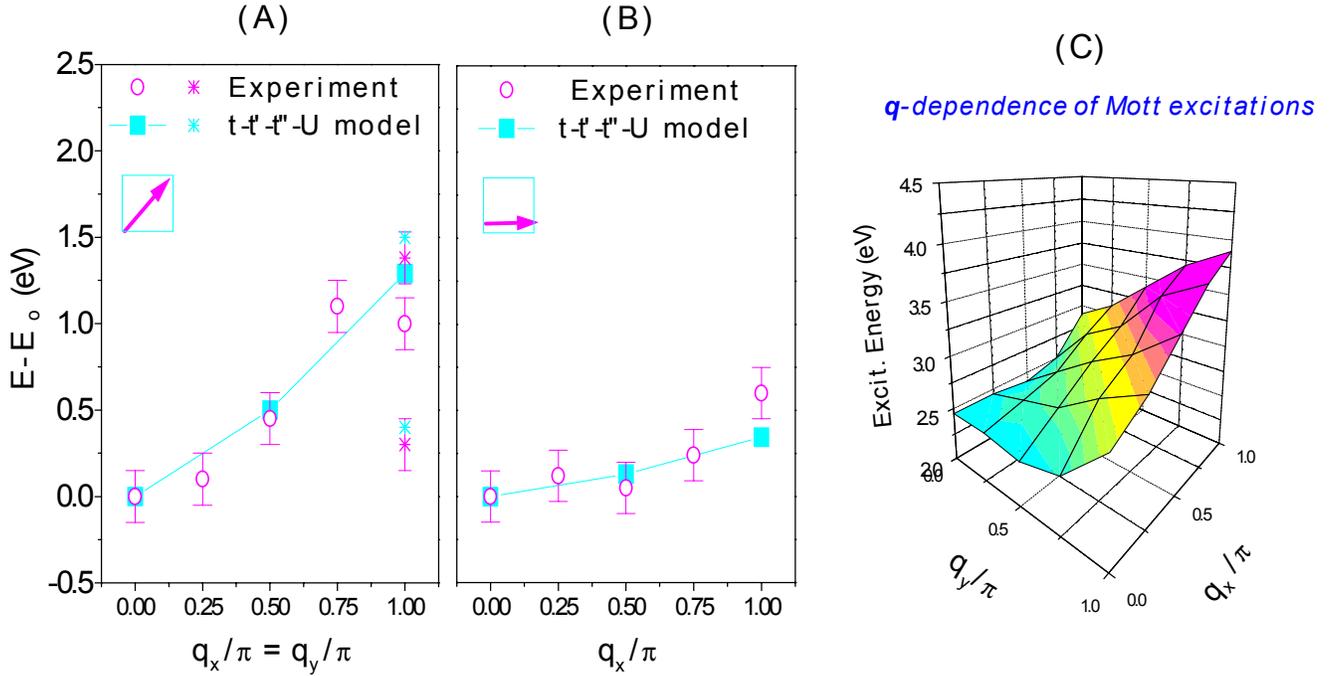

Hasan *et.al.,* Science 288, 1811 (2000)

Fig. 3: (**Dispersion plots**) The experimental and the theoretical results on the momentum dependence of the center of gravity of the low-energy inelastic feature are shown along the <110> direction (A) and the <100> direction (B). Relative excitation energies are plotted referenced to the energy ($E_o$) at **q** = (0,0) along each direction. Open magenta circles and closed cyan squares denote experimental and theoretical centers of gravity, respectively. For **q** = ($\pi,\pi$) (in Fig. (A), fractional **q**-co-ordinate = 1) in addition to plotting the center of gravity (o, magenta) we plot the locations of two separable features denoted by aesteriks for data (magenta) and theory (cyan). Fig. 3C shows a **q**-space map of charge excitations across the Mott gap which summarizes the overall dispersion behavior. (The color in Fig. 3C recodes the energy axis to create a clear visual representation for the shape of the dispersion surface.)

# Fig. 4 ( Mott Excitation Cartoon )

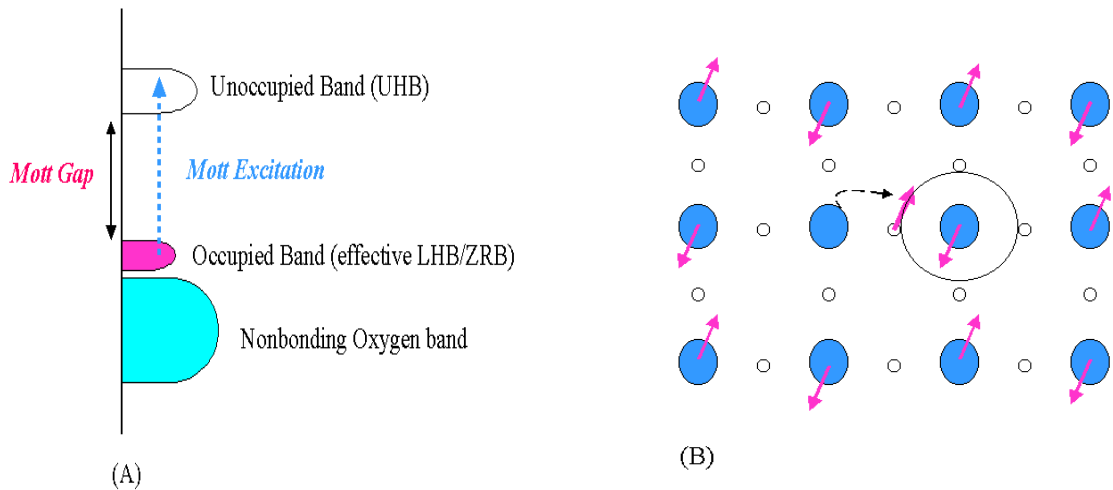

**Hasan *et.al.*, Science 288, 1811 (2000)**

Fig. 4: (**Mott excitation cartoons**) (A) A schematic of (**k**-integrated) electronic structure of the parent copper oxide insulator is shown. There exists a charge excitation gap between the occupied band and the unoccupied upper Hubbard band *(23,28)*. The arrow denotes the excitation of an electron. (B) A schematic of particle-hole pair excitations in a $CuO_2$ square lattice with long-range antiferromagnetic correlation. The arrows denote the spins of holes. The 'filled dark-cyan' and 'empty white' circles denote Cu-sites and O-sites respectively. The (dark-cyan) empty site in the middle is the electron (particle) and the big complex containing four oxygen sites is the Zhang-Rice singlet (hole) *(9)*. The experimental results reveal that the propagation of particle-hole excitations is strongly anisotropic in this antiferromagnetically ordered lattice.